# Colour for behavioural success

Birgitta Dresp-Langley* & Adam Reeves#

*ICube UMR 7357 Centre National de la Recherche Scientifique, University of Strasbourg, Strasbourg, France

#Department of Psychology, Northeastern University, Boston, MA, United States

## Abstract

Colour information not only helps sustain the survival of animal species by guiding sexual selection and foraging behaviour, but also is an important factor in the cultural and technological development of our own species. This is illustrated by examples from the visual arts and from state-of-the-art imaging technology, where the strategic use of colour has become a powerful tool for guiding the planning and execution of interventional procedures. The functional role of colour information in terms of its potential benefits to behavioural success across the species is addressed in the introduction here to clarify why colour perception may have evolved to generate behavioural success. It is argued that evolutionary and environmental pressures influence not only colour trait production in the different species, but also their ability to process and exploit colour information for goal-specific purposes. We then leap straight to the human primate with insight from current research on the facilitating role of colour cues on performance training with precision technology for image-guided surgical planning and intervention. It is shown that local colour cues in 2D images generated by a surgical fisheye camera help individuals become more precise rapidly across a limited number of trial sets in simulator training for specific manual gestures with a tool. The progressively facilitating effect of a local colour cue on performance evolution in a video controlled simulator (pick-and-place) task can be explained in terms of colour-based figure-ground segregation guiding attention towards local image parts, with an immediate cost on performance time, and a lasting long-term benefit in terms of greater task precision.

## Introduction

In order for colour to have a functional meaning, in terms of a driving factor for behavioural success, colour must be detected and encoded by the visual receptors of a living organism, be subsequently processed by its nervous system, and affect action. Such processing enables what is commonly called colour perception. As will be made clear in this review, the ability to detect, process and act in response to colour information may be a direct consequence of evolutionary pressure. It is argued that such ability has evolved to a greater or lesser extent indifferent species, and is determined by the functional anatomy and epigenetic development of their visual systems. At least forty different types of visual system exist in the animal world. The simplest is found in *aplysiae* (e.g. Goelet et al., 1986), a species that has been extensively studied by Eric R. Kandel and his team and only seems able to differentiate light from dark. The visual brains of higher order species are able to discriminate not only light intensities, but also shape and colour. An enormous diversity in animal retinal structures and visual neuronal mechanisms is found, with a corresponding diversity in the functional role of species specific colour vision, perception and behaviour (e.g. Land & Nilsson, 2002, Land et al., 2007). This diversity can be explained by the fact that separate evolutionary processes have acted on the different species. We will first start out with a brief overview of how anatomically and functionally different visual systems have enabled colour vision in different species.

A living organism possessing a single retinal pigment would most likely see the world in monochrome, as we do at night when only our 'rod' photoreceptors are active. In addition to rods, most species have cones, the photoreceptors that provide colour perception during the day. We can define the input to these photoreceptors in physical terms as the overall level of light provided by the sun or moon between about 360nm and 700nm and colour as the distribution of wavelengths within this range (e.g. McLeod & Boynton, 1979). The number of retinal pigments in many animal species is not, as it is in the human (e.g. Bagnara, 1998), limited to three. Of vertebrate species studied so far, the most developed ability to discriminate stimuli on the basis of their colour is found in birds (Andersson & Amundsen, 1997). Birds vary according to species in their capacity to discriminate colour. Diurnal birds tend to have increased ultraviolet (UV) sensitivity, either UVA or UVB (Andersson & Amundsen, 1997; Cuthill et al., 1999; 2000) and have more cones, while nocturnal species have a much higher proportion of rods (Ruggeri et al., 2010). UV wavelength sensitivity improves the temporal resolution of the avian visual system by enabling the detection of rapid flicker movements (Rubene et al., 2010), which may be critically important to ensure their behavioural adjustment when they fly, or have to catch prey. Microspectrophotometry has been employed to test the spectral sensitivity properties of animal retinas. With this technique and through behavioural studies, it has been deduced that four cone classes exist in approximately thirty species of birds. Some even have five types of photoreceptors: four single cone classes and, for about half of all cones, a double cone class with two different photopigments (Bowmaker et al., 1997). In addition, the spectral sensitivities of certain cones are narrowed by a coloured oil droplet filtering light signals (Partridge, 1989; Bowmaker et al., 1997), with the double cone class having a different oil droplet filter and broader spectral tuning (Bowmaker et al., 1997). By sequencing the part of the gene coding for avian retinal opsins, Odeen and Hastad (2003) have shown that the evolution of avian colour vision is more complex than had previously been thought. Their data support the theory that sensitivity biases toward violet (short wavelengths) has evolved independently at least four times, which may explain why some species such as the chicken (Osorio et al., 1999) process colour inputs with at least three opponent mechanisms: one comparing the outputs of ultraviolet- and short-wavelength-sensitive receptors, one comparing the outputs of medium- and long-wavelength receptors, and a third comparing outputs of short- and long- and/or medium-wavelength receptors. Diurnal birds tend to be tetrachromatic rather than trichromatic and are able to detect chromatic variations in skylight that are invisible to us (as we are insensitive to UV), and provide information about updrafts or other atmospheric conditions that are critical to their survival. Another advantage is related to sexual selection, as colour discrimination is enhanced by seeing UV reflections from the wings of potential mates in many birds (e.g. Boulcott et al., 2005), and in particular starlings and zebra finches (Bennett et al., 1996; 1997).

Retinas with four classes of cones involved in colour discrimination (tetrachromatic vision) have also been reported in fish (Palacios et al. 1998). Most of the latter have photoreceptors with peak sensitivities in the ultraviolet range. 95% of all known fish species seem to discriminate red, yellow/green/blue, violet, and UV up to 365nm. However, since sea water selectively absorbs longer wavelengths, i.e. red light, many fish living below 10 metres detect poorly in the red region. Nevertheless many reef fish species living at this depth emit red fluorescence, the origin of which are guanine crystals, and seem able to detect this colour (Michiels et al., 2008). Fluorescence is widespread in marine organisms and was recently also discovered in some terrestrial chameleon species, which have evolved fluorescing bony tubercles protruding from their skull. The fluorescence becomes visible in males only, shining through their scales under UV light at 365 nm (Prötzl, Heβ, Scherz, Schwager, van't Padje and Glaw, 2018). The fluorescent signal is likely to play a role in sexual selection by helping to attract females.

Poralla & Neumeyer (2006) showed in behavioural experiments that there is no abrupt transition between green and red, but that there is yellow in-between. Kelber, Vorobyev & Osorio (2003) more recently have discussed how photoreceptor signals are combined and compared to allow for the discrimination of biologically relevant stimuli. Five retinal pigments have been found in butterflies and many more in sea organisms such as the mantis shrimp and the sea manta, a giant ray (Cronin, Caldwell & Marshall, 2001; Marshall, Cronin & Kleinlogel, 2007). While the human eye cannot detect all of the electromagnetic spectrum emitted by the sun, and UV being in any case absorbed by the human lens and cornea, lobsters, shrimp, goldfish, trout, bees, and lizards seem to be able to detect light in the UV region (e.g. Fleischmann, Loew, and Leal, 1993; Cronin, Caldwell, and Marshall, 2001; Brink et al., 2002). Comparative psychophysics have shown that the goldfish displays the phenomenon of colour constancy (Neumeyer, Dörr, Fritsch, and Kardelky, 2002), previously found only in primates and avian species.

Research data from genetics suggest that the variations in colour vision across species may result from the fact that the capacity to detect colour has evolved more than once, with gene duplication for visual opsin pigments (Bowmaker, 1998). In addition, probably because of their nocturnal behaviour, certain mammals have rod-dominated retinas (see for example Ruggeri et al., 2010) and many selectively lack cones with the red-sensitive pigment and, therefore, display only dichromatic colour vision. The dichromate cat is incapable of discriminating red, with cones only sensitive to blue/indigo and to yellow/green, as found in the ferret (Calderone & Jacobs, 2003). Horses also have only dichromatic vision detecting long wavelengths and the colours based on them (Carroll et al., 2001). A different system has evolved in rats and mice which have excellent night vision, due to a higher number of rods than cones, but see poorly in colour although both are dichromatic. Both rat and mouse cones co-express two photopigments, one sensitive to wavelengths around 510nm and another sensitive to ultraviolet, and their visual systems exploit these differences to enable them to discriminate certain colours (Jacobs & Williams, 2007; Jacobs, Fenwick & Williams 2001; Jacobs, Williams & Fenwick, 2004; Jacobs and Deegan 2nd., 1994). Diurnal rodents and rodents which live in almost lightless conditions have been found to have similar colour vision (Williams, Calderone & Jacobs, 2005; Jacobs et al., 2003). Genetic data have also confirmed the long held suspicion that colour vision in primates including humans, apes, and Old World monkeys, is better developed than in other mammals (Jacobs, 1993). In spite of data lacking on many species, primates are largely considered to be trichromatic, although variations exist (Jacobs, 1993; Jacobs, 1996; Jacobs & Deegan 2nd., 1999). Variations amongst New World monkeys are even greater, some species being trichromatic while others are only dichromatic (Jacobs and Williams, 2006). In addition, evidence predicts that all male New World monkeys are dichromatic while, depending on their opsin gene arrays, individual females can be either dichromatic or trichromatic (Jacobs & Deegan the 2nd, 2003; 2005; Rowe & Jacobs, 2004). Some nocturnal species appear to be monochromatic (Jacobs 1996). The situation for aquatic mammals is quite different as many species, and in particular mammals that live in deep water, are monochromatic and tend to have blue shifted vision compared to that of many terrestrial mammals (Fasick et al., 1998). This is considered to result from the absence of evolutionary pressure to maintain colour in a dark monochromatic oceanic environment (Newman and Robinson, 2005). On the other hand, behavioural data on the Bottlenose Dolphin (*Tursiops Truncatus*), which loves to stick their head out of the water and seems to enjoy daylight, suggest some colour vision based on rods and a single class of cones (Griebel & Schmid, 2002).

According to Darwin (1871), the choice of a sexual partner represents an important step towards the most successful propagation of a given species by a sexual process. Partner choice in animals for sexual selection is therefore a major factor in the survival of the species (Andersson, 1994). The choice of a sexual partner in the animal kingdom is determined in one of three distinct ways. In polygamous animals, the male usually eliminates potential rivals by chasing them away aggressively. After elimination of competitors, the female, and often a harem of females, is left with one alpha male as sexual partner by default. In addition to leaving little freedom of choice to the female of the species, the successful aggressive displays can be viewed by the female as a reflection of the most successful genes (Darwin, 1871). Alternatively, a male "seduces" a potential female by parading his colourful attributes as ornaments, which correspond to his secondary sexual characteristics. Such "nuptial" behaviour frequently leads to exaggerated display of the ornaments, which include shimmering of peacock feathers (Zi et al., 2010) or lifting the bright blue webbed feet of the blue footed boobies of the Galapagos archipelago, well known for having inspired Darwin's theory of evolution during his studies there in 1835. The female thus has a selective choice, based on her estimation of the "best" genes, which may be considered to reflect how impressive her potential partner's ornaments are in terms of size and colour intensity (Price, 2006), signalling the physical condition and immune competence of the male. This process is frequently employed in monogamous animals and in particular in monogamous birds. The female preference to mate with males with exaggerated external ornaments may have arisen from evolutionary pressure (Cornwallis & Uller, 2010), when a female preference for some aspect of male morphology provoked selection for males with the appropriate ornament, a possibility generally known as the *"sexy son" hypothesis* (Weatherhead & Robertson, 1979). Plasticity (Danielson et al., 2006) in sexual mate choice behaviour has been observed in birds during the course of the breeding season, which is accompanied by a reduced variation in the size and brilliance (brightness) of colourful ornaments. Later in the season the choice based on male ornaments is replaced by a choice based on genetic complementation (Badyaev et al., 2001). Although most studies have focused exclusively on the role of female mate preferences in maintaining or promoting colour variation, a recent study (Pryke and Griffith, 2007) highlights that both female and male components of mate choice should be taken into account. Females showed a strong preference for mates with the most elaborate sexually dimorphic traits, while males were particularly choosy, associating and pairing only with females of their own morph-type. Sometimes, even in highly coloured male and female fish (Sköld et al., 2008), the males rather than the females make the choice of a sexual partner (Houde et al., 1990), which suggests that female ornamentation male also be sexually selected (Amundsen and Forsgren, 2001).

In vertebrates such as birds, the success of sexual selection is critically determined by both the successful production of colour traits (Hill & McGraw, 2006), and by their successful visual detection and perceptual processing (Cuthill et al., 1999). Successful reproductive sex in animals constitutes a process of combining and mixing genetic traits, often resulting in the specialization of organisms and to form offspring that inherit traits from both parents (Andersson, 1994). Genetic traits, contained within the DNA of chromosomes, are passed on from one parent to another in this process. Because of their motility, animals often engage in coercive sex. In a large number of essentially monogamous species, which include humans, the selection of a sexual partner is most often mutual. In species which are not monogamous, such as lions for example, male domination is often determinant. Since certain inherited characteristics may be linked to specifically sex-associated chromosomes, physical differences are frequently observed between the different sexes of an organism. These have been particularly well studied in birds, where major differences in feather colours of males and females may be observed (Pryke & Griffith, 2006; 2007). Over recent years an overwhelming bulk of evidence has shown that these differences play a significant role in mate determination, with males frequently displaying these attributes with the aim of attracting a female partner (Hill & McGraw, 2006, for review). Ornamental colours are critical secondary traits in this context, as secondary sexual traits are classically considered to be derived by sexual selection for traits which give an individual an advantage over rivals in courtship or aggressive interactions (Darwin, 1871, Andersson, 1994). The evolution of secondary sexual traits is not a recent phenomenon in evolution (Tomkins et al., 2010), but harks back to *Dimetrodon* and *Edaphosaurus*, the earliest examples of sexual dimorphism known in the animal kingdom.

Both pigment-based colourations and ultraviolet reflectance, which is structural colour and not perceptible by humans, play an important role in communicating critical information to potential mates. The biological significance of structural colour was first investigated mainly in birds (for reviews see Auber, 1957; Dyck, 1976). More recent reports have emphasized the potential biological role of pigment-based colour, structural colour (Vukusic & Sambles, 2003) including UV reflectance, and fluorescence in avian communication (Andersson & Amundsen, 1997; Hunt et al., 1997, 2001; Örnborg et al., 2002; Siitari et al., 2002; for recent reviews see Hill and McGraw, 2006). There is agreement that multilayer reflectors (Ghiradella et al, 1972; Ghiradella, 1994) at the origin of structural colourations in butterflies are found in other species (Parker, 2000). They were present early in evolution and have even been found in beetle fossils, for example (Parker, 1998; Parker and McKenzie, 2003). Coherent scattering is the optical mechanism that explains colour production in all structurally coloured butterfly scales (Prum, Quinn & Torres, 2006). Male individuals of various avian species exhibit conspicuous colours on their feathers evolved by sexual selection driven by mating preferences (Andersson, 1994; Darwin, 1871; Solis, et al., 2008), such as the males of the cooperative breeding azure-winged magpies (*Cyanopica cyanus*), where only a fraction access the breeding status. These birds display conspicuous blue plumage colourations, and all males that become breeders have a more brilliant and more saturated blue colouration, with a more intense violet hue compared with the non-breeders (Solis et al., 2008). The additional capacity of avian ornaments to reflect UV plays an important role during sexual displays (Hausmann et al., 2003; Hunt et al., 2001; Andersson & Amundsen, 1997; Bennett et al., 1997; Finger, Burkhardt and Dyck, 1992; Finger & Burckhardt, 1994; Siitari et al., 2002; Parker, 1998; Bennett, Cuthill, Partridge and Lunau, 1997). Experimental alterations of the UV component in plumage have been shown to significantly affect sexual signals in many bird species (Maier & Bowmaker 1993; Bennett et al. 1996, 1997; Andersson & Amundsen 1997; Hunt et al. 1997, 1998, 1999).

King Penguins (*Aptenodytes patagonicus*) display highly coloured ornaments, notably yellow/orange breast and auricular feathers and the two orange/pink UV reflecting beak horns (Dresp, Jouventin, and Langley, 2005; Dresp & Langley, 2006, Dresp-Langley and Langley, 2010) on each side of the beak. These colourations have been suggested to signal the health status and immunocompetence of individuals, seem to be detected by individuals of both sexes (Nolan, Dobson, Dresp, and Jouventin, 2006), and are therefore likely to ensure the successful mating of the fittest individuals. During courtship, King Penguins flaunt their beak ornaments by raising their heads high in the air when calling out for a potential partner. The iridescence of their beaks is reminiscent of the shimmering iridescence of peacock feathers, which may explain why the King Penguin is also quite often perceived as a rather "sexy beast" by our own species (Carmichael, 2007). The fact that the beak horn also displays an orange-to-pink colour tone increases the signal function, as more than a single type of photoreceptor in the observing individual would be activated. The perception of this colour signal may also be enhanced by a contrast effect, since the tissue surrounding the beak horn is deep black. In addition, the multiplicity of microstructures with slightly different orientations producing the UV reflectance, which was found to obey Braggs law (Bragg & Bragg, 1915; Jouventin et al., 2005; Dresp & Langley, 2006) spreads both the wavelength and also the angle over which it is reflected, thereby producing a more easily perceptible signal. Such UV reflecting ornaments are absent in sexually immature individuals (Jouventin et al., 2005; Massaro, Lloyd & Darby, 2003). Apart from the function of visually attracting potential mates, ornamental colours communicate information about the general fitness of individuals, in particular whether an individual is well nourished and healthy, providing additional criteria particularly for males in attracting female partners (McGraw et al., 2002). In addition to pure survival, mate selection, breeding performance, fecundity and growth, colour traits appear to be associated in various ways with immunity (Hamilton & Zuk, 1982; Hill & Montgomerie, 1994; Svensson, Raberg, Koch, & Hasselquist, 1998; Lochmiller & Deerenberg, 2000; Blount et al., 2003; Nolan, Dobson, Dresp, & Jouventin, 2006).

Another important step in ensuring the survival of a species is foraging (Dall & Johnstone, 2002). After mating and reproductive sex, it becomes necessary to find food sources to ensure the survival of the offspring, and learning to discriminate between the colours of fruits and foliage acquires is a critical function to successful foraging behaviour, especially in birds. Forage success is determined by whether or not a food source is recognized at the right time by individuals of a species. Under conditions of abundance, this does not represent a problem, but in natural environments, food sources are often scarce or irregularly distributed, and their availability varies over time. Thus, the foragers need to identify profitable food sources and to retain information relative to their spatial location to be able to return to them for meeting nutritional requirements. One and the same food source rarely produces the sought after nutrient all year long, and foraging in a dynamic environment therefore requires a balance between exploitation of known resources and the exploratory sampling of new resources (e.g. Dall & Johnstone 2002), two types of foraging behaviours referred to as *"win-stay" and "win-shift" strategies* (Sulikowski & Burke, 2011). Associating a visual cue such as colour with a given food reward increases the rate of discovery of further food sources and thereby helps individuals of a species cope with the spatial heterogeneity of food resource distribution in a natural environment (Edwards et al. 1996). Examples of increased foraging efficiency brought about by visual cues such as colour in a variety of species, including vertebrates (Johnson 1991; Durier & Rivault 2000; Hurly & Healy 1996; Warburton 2003), suggests that animals learn to identify food sources on the basis of such cues. Colour is one of these, and combinations of cues include other visual, olfactory, auditory, tactile, and gustatory ones (Montgomerie & Weatherhead 1997; Hill et al. 2001; Croney et al. 2003; Goyret & Raguso 2006; Ishii & Shimada 2010; Arenas & Farina 2012). The use of colour cues for successful foraging may be particularly important under conditions of temporal and spatial variation in food availability. Pollinators forage on irregularly distributed plants and flowers, and fruit or nectar availability varies over time depending on flowering times. The availability as well as the quality of a given food reward varies between flowers on the same plant, depending on the age of the flowers, the time of day, and other environmental conditions (Corbet et al. 1979). To locate and remember high-quality food sources, birds often use colour cues to direct and to optimize their foraging (Gill and Wolf, 1977). Bird-pollinated flowers are precious food sources for bees. They commonly display bright yellow or red colourations (Rodríguez-Gironés & Santamaría 2004), and there is evidence that angiosperm flowers have evolved chromatic cues that suit the colour discrimination abilities of their pollinators, most likely to reinforce visits by the most efficient of them (Shrestha et al. 2013).

Many animals can form associations between colour and reward (Vallortigara 1996; Kelber 2005; Raine & Chittka 2008). Studies on the foraging behavior of free-living rufous hummingbirds (*Selasphorus rufus*) have shown that these birds retain information relative to high-quality food sources on the basis of learnt colour cues (Hurly & Healy 1996; Hurly & Healy 2002; Henderson et al. 2006; Healy & Hurly, 2013). They predominantly use spatial information when returning to repeatedly sampled flowers (Hurly & Healy 1996; Marshall et al. 2012), and rely on learnt colour cues for foraging in unfamiliar places and the exploratory sampling of new flowers. Goldsmith & Goldsmith (1979) found that hummingbirds could learn to associate colour with nectar reward within less than twenty visits to the food source. They learn rapidly to associate colour with the presence or absence of reward (Goldsmith & Goldsmith 1979; Meléndez-Ackerman et al. 1997; Hurly & Healy 2002), or to associate colour with either food volume or concentration (Bateson et al. 2003; Bacon et al. 2010). Colour cues also enhance their ability to return to target flowers in the presence of distracting flowers (Hurly & Healy 1996). Coexisting species of hummingbirds were found to forage on the basis of learnt colour cues, selecting almost exclusively high-quality food sources and ignoring poor-quality ones (Sandlin, 2000). Other avian species learn to combine spatial cues and colour to detect and recognize rewarding flowers or feeders as, for example, rainbow lorikeets (Sulikowski & Burke 2011). Although the nomadic lifestyle of sunbirds does not facilitate studying their perceptual abilities in the wild, a recent study in semi-natural enclosures (Whitfield, Köhler, and Nicolson, 2014) has shown colour-learning and colour-cue-use abilities in the captive amethyst sunbird (*C. amethystina*). Following training to colour cues, amethyst sunbirds more effectively selected high-quality food sources at the first attempt and throughout the feeding period, thereby decreasing the total time spent on feeding and the number of visits to the food source while increasing the rate of energy gain per meal. These findings are in agreement with earlier studies on foraging behaviour of sunbird species in natural environments (Gill & Wolf 1977). A higher number of visits to abundant food sources in the presence of colour would lead to suggest that the colour cues facilitate the return of sunbirds to rich food sources (Scoble & Clarke (2006) in the natural environment. Sunbirds forage over large distances, on plant species with characteristically red or orange flowers (Johnson and Nicolson 2008) with variable nectar concentrations. The birds rapidly adjust their feeding patterns to maintain an energy balance (Köhler et al. 2008). Sunbirds foraging without colour cues are likely to sample more frequently from a larger number of poor feeders compared with birds that have learnt to use colour cues. As a consequence, the latter return to the food source less frequently and display shorter times spent there (Whitfield, Köhler, and Nicolson, 2014).

It takes a giant leap up the phylogenetic scale, from the avian species to the human primate, to unravel the functional role of colour in the behavioural success of our own species. In humans, the colour quality of hair and skin may to a given extent contribute to whether or not a first date between individuals will be successful. Similarly, the colour of a meat, fruit, or vegetable displayed in a supermarket may influence humans when choosing from several available options, and recent research data have revealed that goods with fully saturated colours are instantly perceived as larger in size than poorly saturated equivalents (Hagtvedt & Brasel, 2017). However, the functional significance of colour for man (Dresp-Langley and Langley, 2010; Pinna & Reeves, 2013) has shifted towards other, more sophisticated aspects of a far more complex process of visual communication. Pinna and Reeves suggest that, in addition to the discrimination of light and surface chromaticity, the visual purpose of color in humans is threefold: to inter-relate each chromatic component of an object, thus favoring the emergence of the whole; to support a part–whole organization in which components reciprocally enhance each other by amodal completion; and, paradoxically, to reveal fragments and hide the whole, as in camouflage — in sum, there is a chromatic parceling-out process of separation, division, and fragmentation of the whole in order to best specify the important figure-ground relations in the scene. The evidence they provide is psychophysical, principally based on illusions such as the watercolor effect, and it remains to be seen to what extent these different facets of colour perception can be verified in animals. Although these aspects may seem remote from what we call evolutionary pressure they are, based on what has been discussed previously, likely grounded in evolution. In fact, colour plays a critical role in the cultural development of human society, and well before the effects of the colour of objects on our perception were investigated more or less systematically by the French philosopher Michel Eugène Chevreul (1839), artists had already begun to use colour selectively and effectively to create novel visual sensations in order to refine their techniques and promote the success of their work. In modern art and design, colour not only plays a critical role in generating moods and emotions, but also communicates information relative to what is supposed to be seen as figure and what as ground in complex planar creations. Finally, colour provides visual cues to object qualities that have the power to directly guide human decision making and action, as will be illustrated here on the example of image-guided surgical planning and intervention.

The Renaissance painters had preferentially resorted to *chiaroscuro* and geometric cues to aerial perspective, using a very limited chromatic range, to create landscape depth and figure-ground effects in their works. In the 19th century, at the dawn of abstract expressionism, painters such as Turner, especially in his later work, effectively used a larger variety of colours in combination with aerial perspective to suggest what should be seen as nearer and what as further away to the observer in a painting. He also most successfully combined chromatic brightness and saturation to express and balance figure and ground, and to visually communicate specific moods, and other *qualia*. Later in the evolution of the visual arts and design, architects and designers like Vasarély most effectively manipulated colour brightness and saturation in combination with planar shape geometry, playing with foreground and background effects to generate powerful and complex 3D shape effects in highly abstract configurations which he baptized with strange names. His work as a visual artist is probably among the most relevant for perceptual science, illustrating how colour, luminance, saturation, and shape can be combined with local variations in 2D geometry to elicit powerful visual effects and sensations of three-dimensional structure in the plane. For some time, chromatic and achromatic pathways in the visual brain were considered independent by the visual neurosciences, mostly on the basis of partial evidence from functional neuronal anatomy (e.g. Page & Crognale, 2005). Yet, Leonardo da Vinci in his *Trattato della Pittura* (1651), and subsequently also Chevreul (1839), had already discussed contrast and colour combinations as powerful cues to pictorial depth and how contextual variations may influence such perceptions. More recently, visual science has confirmed that contrast, colour, and shape effects, together or independently, can determine what will be seen as nearer or further away in planar image configurations (Dresp and Fischer, 2001; Guibal and Dresp, 2004; Dresp-Langley & Reeves, 2012, 2014). Others have shown conditions under which surrounding light, contrasts, shadows, and nearby objects can determine how we perceive the given colour as a function of the context (Devinck, Delahunt, Hardy, Spillmann, and Werner, 2005; 2006; Devinck and Spillmann, 2009; Cao, Yazdanbakhsh, and Mingolla, 2011). Most contemporary visual artists probably would readily agree that, in addition to planar geometry, colour can be a powerful and potentially self-sufficient medium for creating structural effects. Science has only just begun to unravel some of functional contribution of colour to the wider development of human cultural activities as well as modern technology, as in the healthcare domain, where colour is exploited by imaging techniques as a powerful tool for guiding, and thereby promoting the success of, complex human interventions.

The use of colour cues has become an important strategy in recent advances in visual interface technology for image-guided surgical planning and intervention. Image-guided surgery uses images taken before and/or during the procedure to help the surgeon navigate. The goal is to augment the surgeon's capacity for decision making and action during the procedure (see Perrin et al., 2009, and Dresp-Langley, 2015, for reviews). Colour is strategically added to images at various stages of the process. In this kind of augmented reality by colour, guidance is provided directly on the surgeon's view of the patient by mixing real and virtual properties of human tissue and organs. The perceptual quality of the colour tone, brightness, or saturation in the rendered image is essential for making specific regions of interest to the surgeon optimally perceptible. This includes the visual traceability of devices relative to the patient, the registration and alignment of the preoperative model, and optimized rendering and visualization of preoperative data. Visualization by colour in this context means translating image data into a graphic representation that is understandable by the user (the surgeon), and to convey important information for assessing structure and function. In other words, colour is exploited in this context to "make the invisible visible", and for helping make critical decisions as swift and as safe as possible during interventions. Although the field has evolved dramatically in recent years, the most critical problem is still the one of task-centred user interface design, where the timing of the generation of image data is critical, and to facilitate navigation through large cavities with multiple potential obstacles, such as within the human abdomen or the brain. Complex displays designed to provide navigational aids benefit from additional colour cues to visual field depth and tool-alignment. Interventional navigation techniques use surface renderings of anatomy from preoperative imaging with intra-operative visualization techniques, where colour is a highly effective tool for making the invisible visible to the human operator or surgeon. A common strategy here is representing volumetric data as 2D surfaces in colours of varying opacity. The efficiency of renderings in facilitating human perception-based decisions can be evaluated in terms of the perceptual salience of the surfaces that represent the most critical regions of interest for effective and safe human intervention. Intra-operative imaging often provides further diagnostic information and permits assessing risks as well as perspectives of repair.

Image-guided instrument tracking is a major challenge for image-guided surgery technology (West and Maurer, 2004; Huang et al., 2007). In image-guided interventions, the critical problem for the surgeon is eye-hand coordination, which involves keeping track of the relative positions of the surgical tool(s) and of the direction of the tool-movements (e.g., Jiang et al., 2015). The effectiveness of visual displays for tracking tooltips in simulator training is an important factor to skill evolution in trainees (Batmaz, de Mathelin, and Dresp-Langley, 2016 a, b and 2017). In a series of psychophysical experiments designed to study image conditions likely to facilitate simulator training for optimally precise and swift manual gestures with a tool of novice surgeon, we have included the strategic use of a colour cue to test whether it would help novice surgeons to become more precise more rapidly in a computer controlled simulator task guides by a surgical fisheye camera image.

## Material and Methods

The experiments were run on a computer controlled perception-action platform (*EXCALIBUR*) for image-based analysis of data relative to the time (in milliseconds) and precision ( in pixels) of manual operations with a tool in a pick-and-place task. Hard and software components of the simulator platform are described in detail elsewhere (Batmaz, de Mathelin, and Dresp-Langley, 2017, https://doi.org/10.1371/journal.pone.0183789).

### Ethics

The study was conducted in conformity with the Helsinki Declaration relative to scientific experiments on human individuals with the full approval of the ethics board of the corresponding author's host institution (CNRS). All participants were volunteers and had provided written informed consent.

### Subjects

Six surgeons with minimal experience in image-guided interventions, ranging in age between 25 and 35 years, participated in this study. They were all right-handed.

### Objects displayed on the 2D monitor

Video input received from an HD USB fisheye camera was processed by a DELL Precision T5810 model computer equipped with an Intel Xeon CPU E5-1620 with 16 Giga bytes memory (RAM) capacity at 16 bits and an NVidia GForce GTX980 graphics card. Experiments were programmed in Python 2.7 using the Open CV computer vision software library. The computer was connected to a high-resolution color monitor (EIZO LCD 'Color Edge CG275W') with a 30 Hz refresh rate and a 2560x1440 pixel display. The screen has an inbuilt color calibration device (colorimeter) using the Color Navigator 5.4.5 interface for Windows. The video input received by the computer from the HD USB fisheye camera generated the raw image data.  These were adjusted to a viewing frame of 640 pixels (width) x 480 pixels (height) and processed to generate 2D visual displays (Figure 1) in a viewing frame of 1280 pixels (width) x 960 pixels (height), the size of a single pixel on the screen being 0.32mm. Real-world data and visual display data were scaled psychophysically to ensure that the visual display subjectively matched the scale of the Real-world Action Field (RAF) of the simulator as seen in front of the observer as closely as possible. The luminance ($L$) of the light gray (x=3.03, y=3.19, z=3.47 in CIE color space) RAF visualized on the screen was 33.8 cd/m$^2$. The luminance of the medium gray (x=0.66, y=0.70, z=0.76 in CIE color space) target areas was 15.4 cd/m$^2$, producing a target/background luminance contrast (Weber contrast: (($L_{foreground}$-$L_{background}$)/$L_{background}$)) of -0.54. The luminance of the blue (x=0.15, y=0.05, z=0.80 in CIE color space) object surface on the screen was 3.44 cd/m$^2$, when shown in colour and when shown in its achromatic version on the screen. This gave Weber contrasts of -0.90 with regard to the RAF, and -0.78 with regard to the target areas in both achromatic and colour viewing modes. The luminance (29.9 cd/m$^2$) of the green (x=0.20, y=0.70, z=0.10 in CIE color space) tool-tips produced Weber contrasts of -0.11 with regard to the RAF, and 0.94 with regard to the target areas. All luminance values for calculating the object contrasts visualized on the screen were obtained on the basis of standard photometry using an external photometer (Cambridge Research Instruments) and interface software. The luminance of the different objects (object top surface, tool-tips, the square-shaped target areas and the grey action field) displayed on the screen was photometrically identical in the two display conditions (Figure 1), yielding identical local (Michelson) contrasts in both viewing modes. All subjects stated to perceive the contrast intensities as equivalent in the two displays (Figure 1). The pick-and-place simulator task here requires placing the square-shaped 3D object on the centre of each of the different dark grey 2D target squares. Optimal precision in this task involves, as pointed out previously (Dresp-Langley, 2015), segregating the blue/grey square top of the 3D object from the gray 2D squares defining the target areas. Psychophysical figure-ground experiments (Guibal & Dresp, 2004) have shown that achromatic patterns with identical Michelson contrasts produce the same perceptual figure-ground response probabilities, however, when local colour cues were added, stronger figure-ground responses were found. Therefore, we are led to assume that any difference in placing precision from the two different display conditions here can confidently be explained by the presence/absence of the local colour cue in the image.

FIGURE 1

*Experimental procedure*

The experiments were run under conditions of free viewing, with illumination levels that can be assimilated to daylight conditions. The RAF was illuminated by two lamps (40Watt, 6500 K) which were constantly lit during the whole duration of an experiment. Participants were comfortably seated at a distance of approximately 150 cm away from the screen, looking straight-ahead at the image. Seats were adjusted individually in height at the beginning of a session to ensure that the image displayed on the monitor was slightly higher than the individual's eyes when looking straight at the screen, which is a near-optimal viewing position for laparoscopic simulator training. All participants were given a printout of the targets-on-RAF configuration with white straight lines indicating the ideal object trajectory, and the ordered (red numbers) target positions the small blue cube object had to be placed on in a given trial set of the pick-and-place task, always starting from zero, then going to one, to two, to three, to four, to five, and back to position zero. Participants were informed that they would have to position the cube with the tool, a forceps-type instrument, held in their dominant hand "as precisely as possible on the center of each target area of the RAF, as swiftly as possible, and in the right order, as indicated on the printout". Before starting the first trial set, the participant could look at the printout of the RAF with the ideal trajectory steps in the right order for as long as he/she wanted. When an individual felt confident to be able do the task, the experiment was started. Each participant was given a blank warm-up trial set, and then trained in ten repeated trial sets in each of the two viewing conditions (with colour cue and achromatic). The order of conditions was counterbalanced within individual sessions and between subjects. in two separate successive sessions. Each time the small object was placed on a target area, the computer software processing the video data (see Batmaz, de Mathelin, and Dresp-Langley, 2016, 2017, 2018) calculated the total number of pixels separating the object from the ideal on-target-centre location. At the end of a trial set, the cumulated number of "off-target" pixels from the five successive task steps ("place" manoeuvers) and the total time of task completion in milliseconds were written to a labeled excel file.

**Results and Discussion**

Results were analyzed in terms of the means M and their standard errors SEM for dependent variables relative to <u>time in seconds</u> and <u>precision in pixels</u> for with two levels (achromatic *vs* with colour cue) of the *display* factor $D_2$, ten levels of the trial rank factor $T_{10}$, and six subjects. We therefore have a balanced Cartesian experimental design $S_6 x D_2 x T_{10}$ yielding a total of 120 data (observations) for time and precision.

For the dependent variable relative to time, the data analysis returned a shorter average time for the achromatic image condition (M=15.6 seconds, SEM=1.5) compared with the image condition with local colour cue (M=16.9, SEM=1.4). Two-way repeated measures ANOVA signals that the difference in means for time (*d*=1.3) between the two levels of the display factor is statistically significant (F(1, 5) = 8.423; p<.05). Average times decrease systematically as the rank number of the trials in a session increases. Means for time as a function of the rank number of trial sets are shown in Figure 2 (top) for each of the two display conditions. The effect of trial rank (F(9, 45) = 10.24; p<.001) and the interaction between the trial rank factor and the display factor on time (F(9,45) = 11.41; p<.001) are statistically significant. The condition of equality of variance of the distributions for 'time' was satisfied.

FIGURE 2

For the dependent variable relative to precision, the data analysis returned a higher average off-target score, indicating lesser precision, for the achromatic image (M=1682, SEM=64) condition compared with the image condition with the local colour cue (M=1396, SEM=59), which produced greater precision. Two-way repeated measures ANOVA signals that the difference in means for precision (*d*=286) between the two levels of the display factor is statistically significant (F(1, 5) = 22.60; p<.01). As the rank number of the trials in a session increases, average precision increases significantly (F(9, 45) = 3.500; p<.01) as testified by the decreasing off-target scores, depending on the display condition shown in Figure 2 (bottom). Average off-target scores get smaller as the trial increases only in the image condition with the colour cue. The interaction between the trial rank factor and the display factor on the dependent variable for precision is statistically significant (F(9, 45) = 9.878; p<.001). The condition of equality of variance of the distributions for 'precision' was satisfied.

Beyond considering the population statistics from analysis of variance, there is a complementary and quite revealing way of looking at the average data for time and precision from this experiment. Trend analysis (linear regression analysis) on the average data reveals the correlation coefficients $R^2$ from a least-squares fitting process of the variance of the residuals to the variance of the one-sample dependent variable, representing 1 *minus* the *ratio* of that variance. $R^2$ gives an estimate of the fraction of variance in the average data that is explained by the fitted trend line. The statistical significance of the trend is determined by the statistical probability that the linear trend is different from zero using the Student distribution (*t*) with *n*-2 degrees of freedom (DF).

Here (see Figure 2), it is shown that the achromatic image conditions produce horizontal (flat) regression lines indicating an absence of any linear trend indicative of a training effect. In other words, the average performance scores for time and precision do not evolve in training sessions with ten trial sets when guided by images without a colour cue. On the other hand, the image condition with the colour cue, on average, produces such training effects, for both time and precision. These effects are revealed by the significant linear trends for this image condition, shown in Figure 2 (top and bottom respectively). Under image guidance with the colour cue, average times and average off-target scores present a significantly decreasing linear trend ((t(1,8) = 17.79; p<.001) and (t(1,8) = 13.27; p<001) as the rank number of the trial set increases. In other words, subjects are progressively getting faster and more precise in ten trial sets when guided by the image with the local colour cue. The one-sample t statistic from the linear trend analysis gives the statistical probability that the trend is real, i.e. significantly different from zero.

Earlier work by Batmaz et al. (2016, 2017, 2018) had shown that the evolution of the performance scores of complete novices and more experienced surgeons in simulator tasks as this one here depend on factors relative to individual speed-precision strategy and on the capacity to select information relative to local detail in the visual display guiding the motor performance. Others (e.g. Yantis & Jones, 1991) had found some time ago that colour can act as a powerful "priority tag" in guiding attention for cognitive performance. The results from this study here show that a local colour cue in the image consistently promotes positive training effects on time and precision in a total of ten trial sets, while an image without the colour cue (achromatic viewing) fails to do so. We suggest that this can be explained in terms of a colour-facilitated figure-ground segregation in images of complex natural scenes with more than two layers of subjective surface depth.

FIGURE 3

This interpretation is consistent with data from Dresp-Langley and Reeves (2012, 2014) who found effects of relative brightness contrast and subject depth induced locally by coloured patches in complex configurations (Figure 3) were significantly stronger in images with more than two subjective surface depth levels in the plane when compared with configurations with only two levels of surface depth in the plane, i.e. only two alternatives for figure and ground. The colour cue most likely guides conscious attention to the manipulated object (e.g. Yantis and Jones, 1991; Dresp-Langley and Durup, 2009; Spillman, Dresp-Langley, and Tseng, 2015), with an immediate cost on task execution times, and a long-term benefit on precision as the training sessions progress further.

**Conclusions**

Evolutionary and environmental pressures influence not only colour trait production in the different species, but also their perceptual ability to see and process colour in terms of useful information for guiding action. This ability directly ensures the success of survival relevant actions directly related to mating (*sexual selection*) and feeding (*foraging*). Moreover, colour not only helps sustain the survival of animal species, but also acts as a key factor in the cultural and technological development of our own species. In the contemporary imaging sciences, the strategic use of colour has become a powerful tool for guiding the planning and execution of interventional procedures (*medical precision technology*). Here we have shown a facilitating effect of a colour cue on training for task precision in a computer controlled simulator (pick-and-place) task. This effect can be explained in terms of a colour-cue based enhancement of figure-ground salience in images with more than two layers of subjective surface depth such as natural surgical images. Colour thus plays an important role in the cognitive development of man by facilitating his adaptation to new and challenging visual worlds in the context of rapidly progressing technological development.

**Acknowledgments**

The contribution of Anil Ufuk BATMAZ for programming the experiments and running the subjects is gratefully acknowledged.

**Figures with captions**

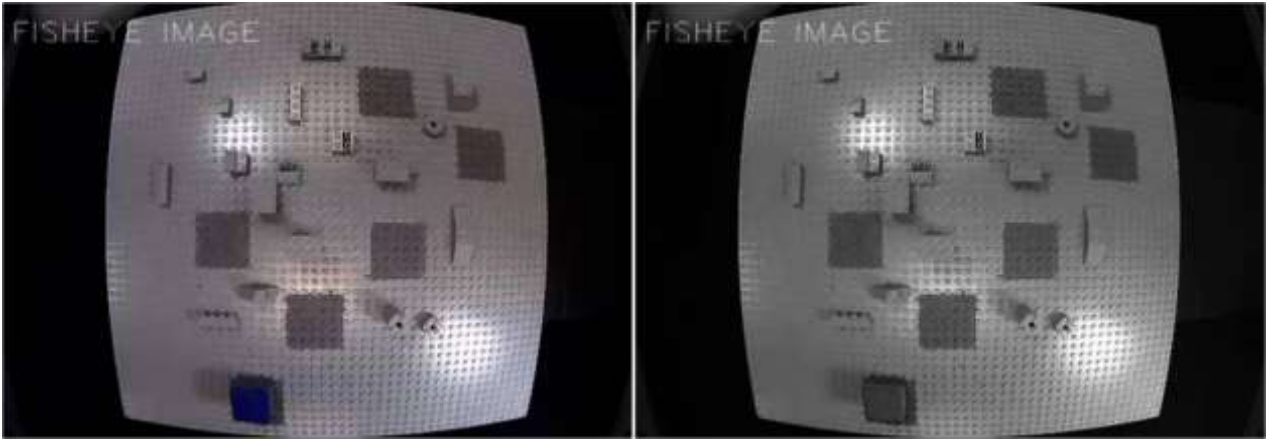

Figure 1: Screen snapshots of the two viewing conditions for image-guided task execution (five step pick-and-place simulator task controlled by a HD fisheye micro camera equivalent to a surgical camera and connected to a computer with optimal processing capacity) with local colour cueing (left) and without (right).

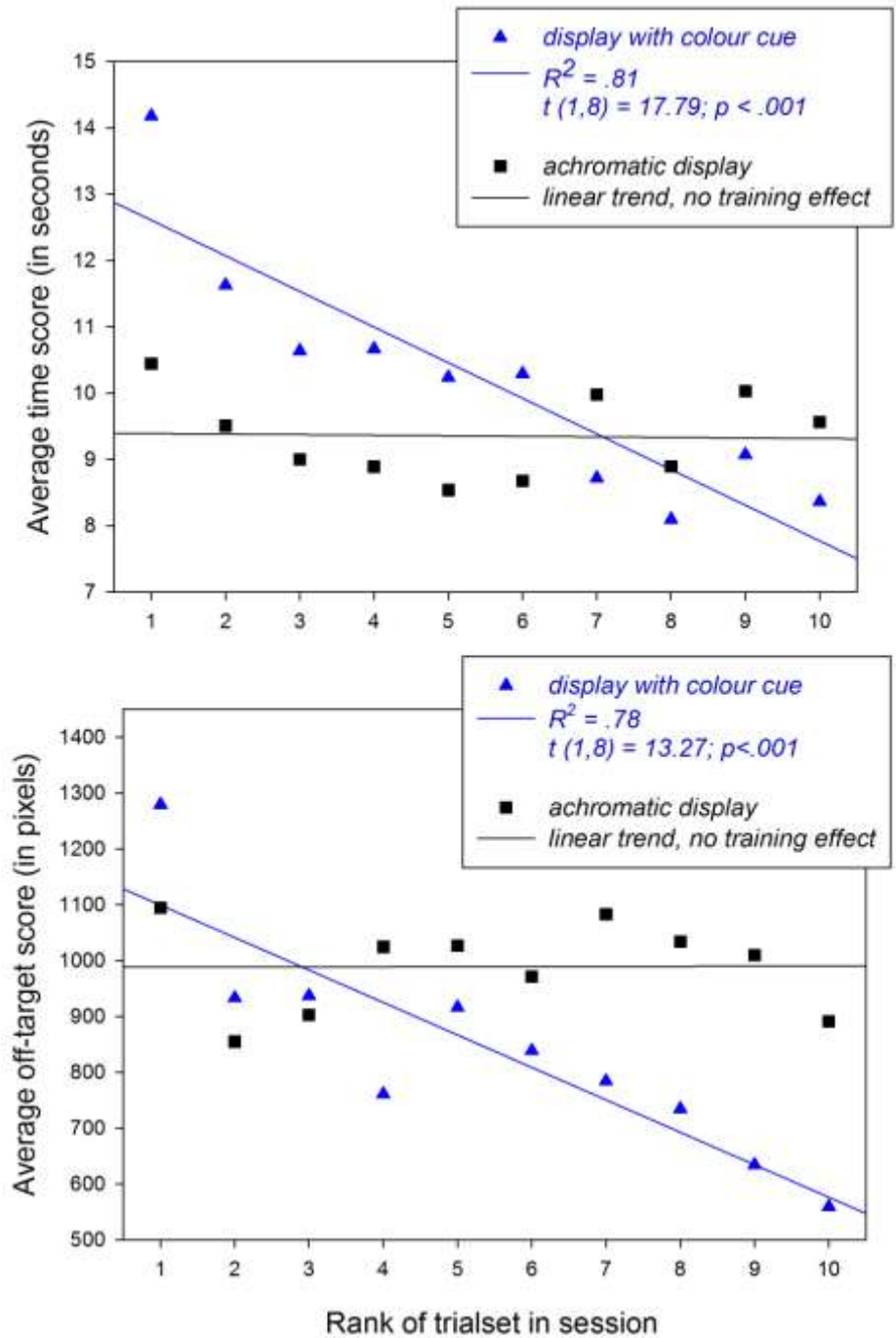

Figure 2: Average data relative to task execution times (top) in seconds, and task precision (bottom) in terms of the number of pixels recorded "off-target-centre" for successive "place" steps of the pick-and-place task.

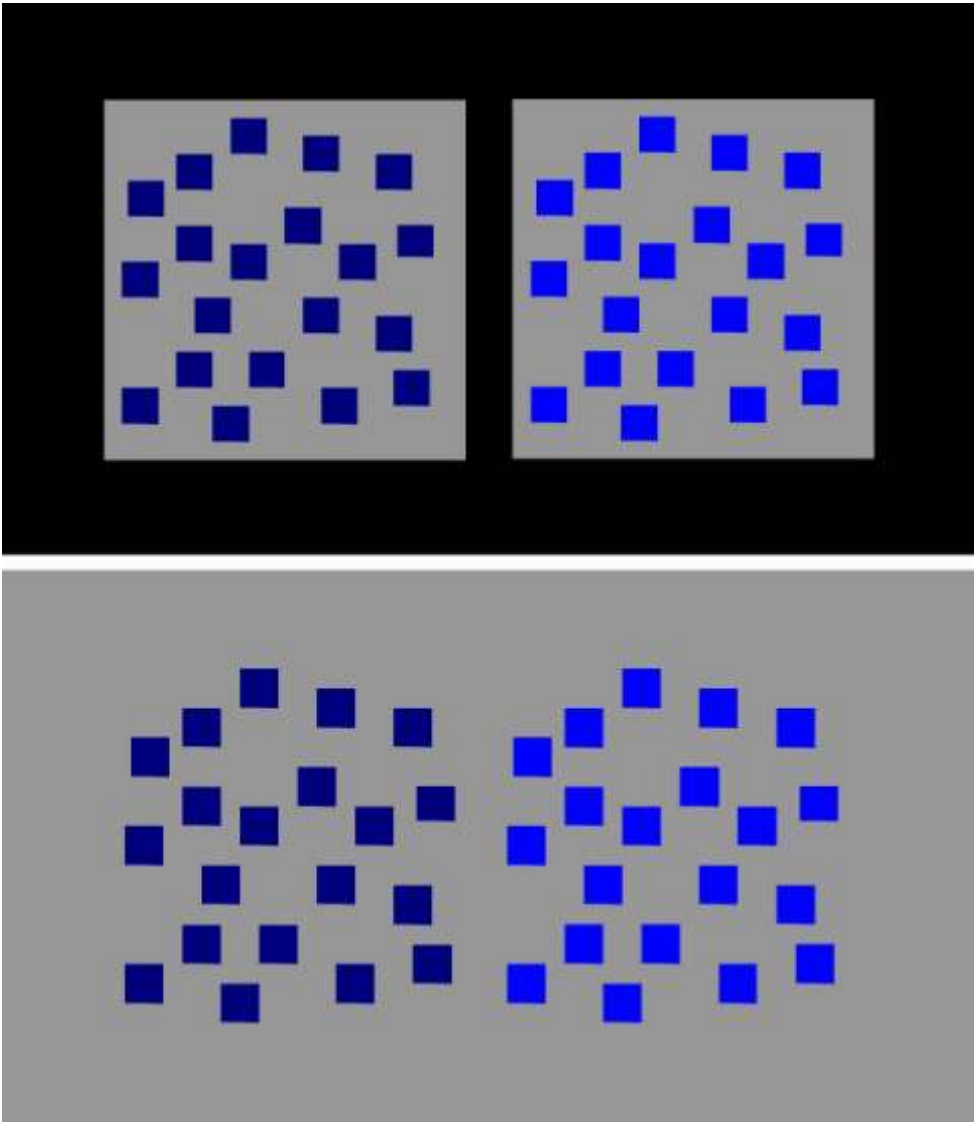

Figure 3: Examples of configurations from Dresp-Langley and Reeves (2012, 2014). In images with more than two subjective surface depth levels in the plane (top), as is the case in most natural scene images including surgical images, effects of relative brightness contrast and subject depth induced by the coloured patches were found to be significantly stronger compared with the same configurations with only two levels of surface depth in the plane, i.e. only two alternatives for what is likely to be figure or ground.